\documentclass[preprint,12pt]{elsarticle}

\usepackage{amssymb}
\journal{ISRN Astronomy and Astrophysics}

\begin{document}

\begin{frontmatter}

%% Title, authors and addresses

%% use the tnoteref command within \title for footnotes;
%% use the tnotetext command for the associated footnote;
%% use the fnref command within \author or \address for footnotes;
%% use the fntext command for the associated footnote;
%% use the corref command within \author for corresponding author footnotes;
%% use the cortext command for the associated footnote;
%% use the ead command for the email address,
%% and the form \ead[url] for the home page:
%%
%% \title{Title\tnoteref{label1}}
%% \tnotetext[label1]{}
%% \author{Name\corref{cor1}\fnref{label2}}
%% \ead{email address}
%% \ead[url]{home page}
%% \fntext[label2]{}
%% \cortext[cor1]{}
%% \address{Address\fnref{label3}}
%% \fntext[label3]{}

\title{The dark matter halo density profile, spiral arm morphology and supermassive black hole mass of M33}

%% use optional labels to link authors explicitly to addresses:
%% \author[label1,label2]{<author name>}
%% \address[label1]{<address>}
%% \address[label2]{<address>}

\author[1,2]{Marc S.\ Seigar}

\address[1]{Department of Physics \& Astronomy, University of Arkansas at Little Rock, 2801 S.\ University Avenue, Little Rock, AR 72204-1099, USA}

\address[2]{Arkansas Center for Space and Planetary Sciences, 202 Old Museum Building, University of Arkansas, Fayetteville, AR 72701, USA}

\begin{abstract}
%% Text of abstract

In this paper, we investigate the dark matter halo density profile of M33.  We find that the H{\tt I} rotation curve of M33 is best described by a NFW dark matter halo density profile model, with a halo concentration of $c_{\rm vir}=4.0\pm1.0$ and a virial mass of $M_{\rm vir}=(2.2\pm0.1)\times10^{11}$ M$_{\odot}$. We go on to use the NFW concentration ($c_{\rm vir}$ )of M33, along with the values derived for other galaxies (as found in the literature), to show that $c_{\rm vir}$ correlates with both spiral arm pitch angle and supermassive black hole mass.

\end{abstract}

\begin{keyword}
%% keywords here, in the form: keyword \sep keyword

%% MSC codes here, in the form: \MSC code \sep code
%% or \MSC[2008] code \sep code (2000 is the default)

galaxies: fundamental parameters --- galaxies: haloes --- galaxies: individual: M33 --- galaxies: spiral --- galaxies:structure --- dark matter.

\end{keyword}

\end{frontmatter}

%%
%% Start line numbering here if you want
%%
% \linenumbers

%% main text

\section{Introduction}

The currently favored cosmological model, Lambda$+$ Cold Dark Matter 
($\Lambda$CDM), is remarkably successful at reproducing the large-scale 
structure of the Universe (Blumenthal et al.\ 1984; Springel et al.\ 2005).
However, small-scale observations have proven harder to explain.  
High-resolution N-body simulations of $\Lambda$CDM structure formation predict 
that the central density profiles of dark matter halos should rise steeply at 
small radii, $\rho(r) \propto r^{-\gamma}$, with $\gamma\simeq 1 - 1.5$ 
(Navarro, Frenk \& White 1997, henceforth NFW; Navarro et al.\ 2004; Diemand 
et al.\ 2005).  
Observations of rotation curves of late-type disk galaxies and dwarf galaxies, 
on the other hand, have shown that quite often, mass distributions with lower 
than predicted densities or with constant density cores, where 
$\gamma\simeq 0$ (i.e., a pseudo-isothermal profile), are preferred (Swaters et al.\ 2003; Gentile et al.\ 2004, 2005; Simon et al.\ 2005; Kuzio de Naray et 
al.\ 2006, 2008; Shankar et al.\ 2006; Spano et al.\ 2008).  This is known 
as the cusp/core problem.  One possibility is that these observations are 
pointing to a real problem with $\Lambda$CDM cosmology, perhaps indicating 
that the dark matter is not cold, but rather warm (Zentner \& Bullock 2002), 
in which case it is easier to produce constant density cores at the centers 
of dark matter halos.  Another possibility is that these late-type galaxies 
have constant density cores because of their late formation (Wechsler et al.\ 
2002) and that earlier-type bulge-dominated galaxies (which form at earlier 
times) will tend to conform to the standard expectations of the theory.  This 
is because the central mass densities of galaxies tend to reflect the density 
of the Universe at their formation time (Wechsler et al.\ 2002).

In this paper we have chosen to model the H{\tt I} rotation curve of M33 from 
Crobelli \& Salucci (2000).  Due to its proximity, M33 can be studied in 
exquisite detail, and it therefore provides a crucial testing ground of our 
ideas of galaxy formation.  Its Hubble classification is SA(s)cd (de 
Vaucouleurs et al.\ 1991), meaning that is of particularly late-type, with 
little or no bulge.  This is reflected in the central supermassive black hole
mass of $M_{\rm BH}<1500$ M$_{\odot}$ (Gebhardt et al.\ 2001), and black hole 
masses tend to be related to the central bulge mass (Magorrian et al.\ 1998; 
H\"aring \& Rix 2004).  In this paper we model the rotation curve of M33 with 
both a pseudo-isothermal profile dark matter halo density model and an NFW 
dark matter halo density model.  We then use parameters derived from these 
fits to look at relations between the dark matter halo and other galaxy 
properties, such as supermassive black hole mass and spiral arm pitch angle.

This paper is organized as follows.  Section 2 describes the observed data and
data analysis.  Section 3 describes how the rotation curve is modeled and how
we derive the baryonic and dark matter halo contributions to the rotation 
curve.  Section 4 discusses our results and Section 5 summarizes our findings.
Throughout this paper, we assume a flat $\Lambda$CDM cosmology with 
$\Omega_m=0.27$ and a Hubble constant $H_0 = 75$ km s$^{-1}$ Mpc$^{-1}$.

\section{Observations and Data Reduction}

We have made use of the {\em Spitzer}/IRAC 3.6-$\mu$m image of M33. The
IRAC observations were taken as part of the Gehrz Guaranteed Time Observer
Program ID 5.  The mapping sequence for each epoch consisted of $\simeq 148$
positions per channel.  Each position was observed with three 12 s frames
dithered with the standard, small, cycling pattern. The FWHM of the 
point-spread function (PSF) at 3.6-$\mu$m is 1.7$^{\prime\prime}$ or 6.9 pc
at the distance of M33. The final mosaic spans an area of $\sim1.^{\prime}0\times
1.^{\prime}2$.  We adopt a distance to M33 of $d=840$ kpc (e.g., Magrini,
Corbelli \& Galli 2007), and
it has a redshift of $z=-0.000597$ (de Vaucouleurs et al.\ 1991).

For dynamical measurements, we make use of the H{\tt I} rotation curve of Corbelli \& Salucci (2000).  We also make use of the inclination corrected H{\tt I} linewidth from HyperLeda\footnote{http://leda.univ-lyon.fr/} of $100.4\pm3.0$ km s$^{-1}$ (e.g., Paturel et al.\ 2003).

For the determination of the spiral arm morphology we have made use of an $R$
band image from the Digital Sky Survey (DSS).

\subsection{Measurement of spiral arm pitch angle}
\label{mpitch}

Spiral arm pitch angles are measured using a two-dimensional fast Fourier
decomposition technique, which employs a program described in Schr\"oder et
al. (1994). Logarithmic spirals are assumed in the
decomposition. 

The amplitude of each Fourier component is given by
\begin{equation}
A(m,p)=\frac{\sum_{i=1}^{I}\sum_{j=1}^{J}I_{ij}(\ln{r},\theta)\exp{-[i(m\theta _p\ln{r})]}}{\sum_{i=1}^{I}\sum_{j=1}^{J}I_{ij}(\ln{r},\theta)}
\end{equation}
where $r$ and $\theta$ are polar coordinates, $I(\ln{r},\theta)$ is the 
intensity at position $(\ln{r},\theta)$, $m$ represents the number of arms
or modes, and $p$ is the variable associated with the pitch angle $P$, defined
by $P=-(m/p_{max})$. Throughout this work we measure the pitch angle $P$ of the
$m=2$ component.

Pitch angles are
determined from peaks in the Fourier spectra, as this is the most powerful
method to find periodicity in a distribution (Consid\`ere \& Athanassoula 1988;
Garcia-Gomez \& Athanassoula 1993).

The image was first projected to face-on. Mean uncertainties of position
angle and inclination as a function of inclination were discussed by 
Consid\`ere \& Athanassoula (1988). For a galaxy with low inclination, there
are clearly greater uncertainties in assigning both a position angle and an
accurate inclination. These uncertainties are discussed by Block et al.\ (1999)
and Seigar et al.\ (2005, 2006), who took a galaxy with low inclination 
($<30^{\circ}$) and one with high inclination ($>60^{\circ}$) and varied the
inclination angle used in the correction to face-on. They found that for the
galaxy with low inclination, the measured pitch angle remained the same. M33
has a relatively low inclination of $\sim 30^{\circ}$, and so the uncertainty
in the inclination angle in this case, does not result in a large error in
the pitch angle we measure for M33.
Our deprojection method assumes that spiral galaxy
disks are intrinsically circular in nature.

\section{Mass modeling}

\subsection{The baryonic contribution}

Our goal is to determine a mass model for M33 from direct
fitting of mass models to its rotation curve.  We perform a bulge-disk
decomposition in order to estimate the baryonic contribution.  We then
determine several different models and try to recreate the nuclear spiral
by minimizing reduced-$\chi^2$.

We first extract the surface brightness of M33 using the {\em Spitzer} 
3.6-$\mu$m image and the IRAF {\tt ELLIPSE} routine, which fits ellipses
to an image using an iterative method described by Jedrzejewski (1987).  In
order to mask out foreground stars, SE{\tt XTRACTOR} (Bertin \& Arnouts 1996)
was used.  An inclination correction was then applied to the surface brightness
profile (de Jong 1996; Seigar \& James 1998) as follows
\begin{equation}
\mu_i = \mu -2.5 C \log {\left(\frac{a}{b}\right)}
\end{equation}
where $\mu_i$ is the surface brightness when viewed at some inclination $i$, 
$\mu$ is the corrected surface brightness, $a$ is the major axis, $b$ is the
minor axis and $C$ is a factor dependent on whether the galaxy is optically
thick or thin; if $C=1$ then the galaxy is optically thin; if $C=0$ then the
galaxy is optically thick (e.g., Seigar \& James 1998; de Jong 1996).  
Graham (2001a)
showed that $C=0.91$ is a good value to use for the near-infrared $K_s$ band.
Adopting a simple reddening law, where extinction falls as the square of
wavelength, it can be shown that a value of $C=0.97$ is appropriate at
3.6-$\mu$m (Seigar, Barth \& Bullock 2008a) and we adopt this value here.

\begin{figure}
\includegraphics{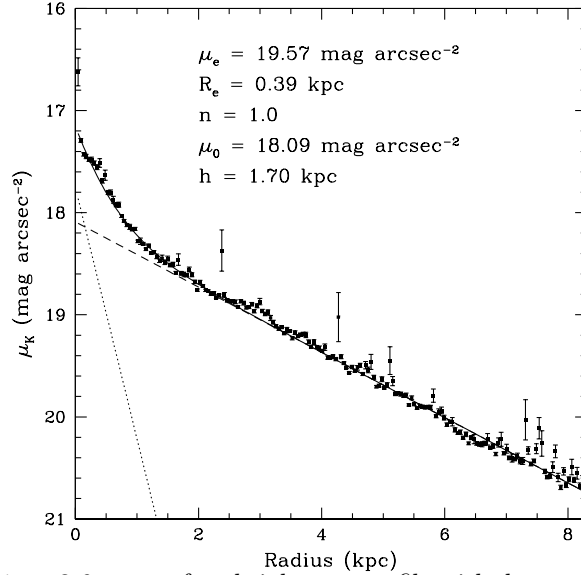}
\vspace*{7cm}
\caption{The {\em Spitzer} 3.6-$\mu$m surface brightness profile with decomposition into bulge and disk components.  The bulge has been fitted with a S\'ersic model (short-dashed line) and the disk has been fitted with an exponential model (long-dashed line).}
\label{SB}
\end{figure}

\begin{table}
\caption{M33 Observational data.  The Hubble type is from de Vaucouleurs et al.\ (1991).  The distance in kpc is taken from Magrini et al.\ (2007).}
\begin{footnotesize}
\begin{tabular}{ll}
\hline
Parameter     &     Measurement \\
\hline
Hubble Type               &     SA(s)cd                \\
Distance (kpc)            &     840                    \\
Position angle of major axis ($^{\circ}$) & 23         \\
Bulge effective radius, $R_e$ (arcmin)    & 1.60$\pm$0.11       \\
Bulge effective radius, $R_e$ (kpc)       & 0.39$\pm$0.03       \\
Bulge surface brightness at the effective radius, $\mu_e$ (3.6 $\mu$m-mag arcsec$^{-2}$)  &  19.57$\pm$0.98 \\
Bulge S\'ersic index, $n$  &        1.0 \\
Disk central surface brightness, $\mu_0$ (3.6 $\mu$m-mag arcsec$^{-2}$) &  18.08$\pm$1.02 \\
Disk scalelength, $h$ (arcmin)     &  6.95$\pm$0.49 \\
Disk scalelength, $h$ (kpc)        & 1.70$\pm$0.12   \\
Disk luminosity, $L_{disk}$ (L$_{\odot}$) & $(3.16\pm0.30)\times10^{9}$  \\
Bulge-to-disk ratio, $B/D$      & 0.03 \\
\hline
\end{tabular}
\end{footnotesize}
\end{table}

The resulting surface brighntess profile {Fig.\ \ref{SB} reaches a surface
brightness of $\mu_{3.6}\sim20.7$ mag arcsec$^{-2}$ at a radius of 
$\sim$13.2 kpc
(equivalent to 54.0 arcmin).  From this surface brightness profile, we 
perform a one-dimensional bulge-disk decomposition, which employs the 
S\'ersic model for the bulge component and an exponential law for the disk
component (e.g., Andredakis, Peletier \& Balcells 1995; Seigar \& James 1998;
Khosroshahi, Wadadekar \& Kembhavi 2000; D'Onofrio 2001; Graham 2001b;
M\"ollenhoff \& Heidt 2001; see Graham \& Driver 2005 for a review). The S\'ersic
(1963, 1968) $R^{1/n}$ model is most commonly expressed as a surface brightness
profile, such that
\begin{equation}
\mu(R)=\mu_e \exp{\left(-b_n\left[\left(\frac{R}{R_e}\right)^{1/n}-1\right]\right)},
\end{equation}
where $\mu_e$ is the surface brightness at the effective radius $R_e$ that
encloses half of the total light from the model (Ciotti 1991; Caon, Capaccioli
\& D'Onofrio 1993). The constant $b_n$ is defined in terms of the parameter
$n$, which described the overall shape of the light profile.  When $n=4$,
the S\'ersic model is equivalent to a de Vaucouleurs (1948, 1959) $R^{1/4}$
model and when $n=1$ it is equivalent to an exponential model.  The parameter
$b_n$ has been approximated by $b_n = 1.9992n - 0.3271$, for $0.5<n<10$
(Capaccioli 1989; Prugniel \& Simien 1997).  The exponential model for the
disk surface brightness profile can be written as follows
\begin{equation}
\mu(R) = \mu_0 \exp{(-R/h)}
\end{equation}
where $\mu_0$ is the disk central surface brightness and $h$ is the disk
exponential scalelength.  The results of our surface brightness fitting are
summarized in Table 1.

We now assign masses to the disk and bulge of M33.  The stellar mass-to-light
ratio in the $K_s$ band is a well-calibrated quantity (Bell et al.\ 2003) which
depends on $B-R$ color.  Seigar et al.\ (2008a) extended this to a 3.6-$\mu$m
image of M31 using the population synthesis codes of Bruzual \& Charlot (2003)
and Maraston (2005).  Using their results, we find a central mass-to-light
ratio of $M/L_{3.6}\simeq 1.25\pm0.10$ with a gradient of -0.014 kpc$^{-1}$.
This results in a disk mass of $M_{\rm disk}=(3.81\pm0.47)\times10^{9}$ 
M$_{\odot}$ and a bulge mass of $M_{\rm bulge}=(1.14\pm0.14)\times10^{8}$ 
M$_{\odot}$ for M33.

A concern in using the 3.6-$\mu$m {\em Spitzer} waveband to determine
the underlying stellar mass, is the effect of emission from hot dust
in this waveband, although this is probably only important in or
near HII regions. In order to place some constraint on this, we have
chosen to explore the emission from dust in the near-infrared $K$
band at 2.2 $\mu$m. Using near-infrared spectroscopy at 2.2 $\mu$m, it has
been shown that hot dust can account for up to 30 per cent of the
continuum light observed at this wavelength in areas of active star
formation, i.e., spiral arms (James \& Seigar 1999). When averaged
over the entire disk of a galaxy, this reduces to a $2$ percent effect,
if one assumes that spiral arms can be up to 12$^{\circ}$ in width. At 3.6
$\mu$m,
this would therefore result in 3 percent of emitted light from dust.

Another concern for the 3.6-$\mu$m waveband would be the contribution
from the polycyclic aromatic hydrocarbon (PAH) emission
feature at 3.3 $\mu$m. However, an Infrared Space Observatory (ISO) 
spectroscopic survey of actively star-forming galaxies by
Helou et al. (2000) found that the 3.3-$\mu$m feature was very
weak when they analysed the average 2.5--11.6-$\mu$m spectrum of 45
galaxies. The contribution of the PAH feature to the 3.6-$\mu$m Spitzer
waveband, is therefore not a major concern.

One other important contribution to the baryonic mass of M33 is the gas mass.
Corbelli \& Salucci (2000) have shown that beyond a radius of 10 kpc, the gas
contributes about the same to the rotation curve as the stars.  Since, the
best current estimate of the gas distribution comes from Corbelli \& Salucci
(2000), we have chosen to adopt their model for the distribution of gas mass
in M33.

\subsection{The dark halo contribution}

\begin{figure}
\label{totrot}
\includegraphics{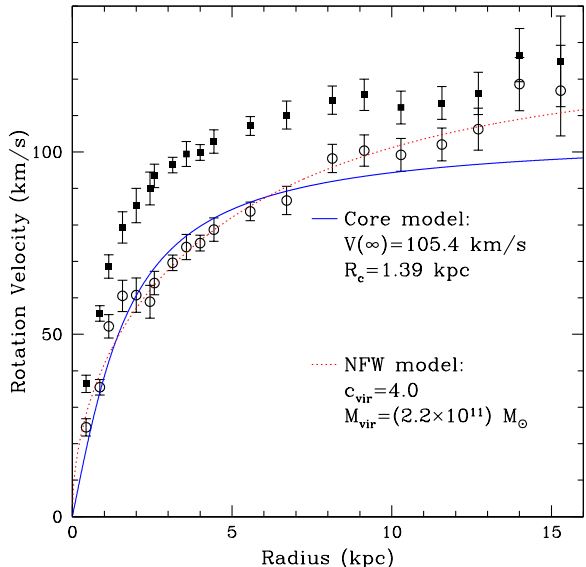}
\vspace*{7cm}
\caption{The H{\tt I} rotation curve from Corbelli \& Salucci (2000) modeled using a pseudo-isothermal model (core model; blue solid line) and a NFW model (red dotted line).  The squares represent the total rotation velocities, and the circles represent the contribution of the dark matter to the rotation velocities (after subtraction of the stellar and gas mass).}
\end{figure}

A range of allowed dark matter halo masses and density profiles is
now explored, using two models for dark matter halo density profiles, the 
pseudo-isothermal model (e.g., Simon et al.\ 2005; see Equation \ref{pseudoi}) 
and the Navarro, Frenk \& White (1997; hereafter NFW) profile.  A pseudo-isothermal density profile is given by
\begin{equation}
\label{pseudoi}
\rho(R)=\rho_{0} \frac{R_{c}^{2}}{R_{c}^{2}+R^{2}},
\end{equation}
which in terms of rotational velocity becomes
\begin{equation}
V_{c}^{2}(R)=V_{c}^{2}(\infty)\left(1-\frac{R_c}{R}\tan^{-1}\frac{R}{R_c}\right),
\label{Vmod}
\end{equation}
where $R_c$ is the core radius, and 
$\rho_0=V_{c}^{2}(\infty)/4\pi G R_{c}^{2}$.
The NFW profile is given by 
\begin{equation}
\label{nfweqn}
\rho(R)=\frac{\delta_c \rho_c^0}{(R/R_s)(1+R/R_s)^2}
\end{equation}
where $R_s$ is a characteristic `inner' radius, and $\rho_c^0$ is the present
critical density and $\delta_c$ a characteristic overdensity.
This overdensity is defined as
\begin{equation}
\delta_c=\frac{100c_{\rm vir}^{3}}{3}
\end{equation}
where $c_{\rm vir}$=$R_{\rm vir}/R_s$ is the concentration parameter and
\begin{equation}
g(c_{\rm vir})=\frac{1}{\ln{(1+c_{\rm vir})}-c_{\rm vir}/(1+c_{\rm vir})}.
\end{equation}
The circular velocity associated with this density is given by Battaglia et 
al.\ (2005) and is
\begin{equation}
V_{c}^{2}=\frac{V_{\rm vir}^{2}g(c_{\rm vir})}{s}\left[\ln{(1+c_{\rm vir}s)}-\frac{c_{\rm vir}s}{1+c_{\rm vir}s}\right]
\end{equation}
where $V_{\rm vir}$ is the circular velocity at the virial radius $R_{\rm vir}$ and $s=R/R_{\rm vir}$.
This NFW profile is a two parameter function and completely
specified by choosing two independent parameters, e.g., the virial
mass $M_{\rm vir}$ (or virial radius $R_{\rm vir}$) and concentration 
$c_{\rm vir}=R_{\rm vir}/R_s$ (see
Bullock et al. 2001a for a discussion). Similarly, given a virial mass
$M_{\rm vir}$ and the dark matter circular velocity at any radius, the halo
concentration $c_{\rm vir}$ is completely determined.

We now proceed by finding the best-fitting NFW and pseudo-isothermal (or 
constant density core) dark 
matter halo density profiles that describe the complete H{\tt I}
rotation curve of 
M33 as observed by Corbelli \& Salucci (2000).   The result
of this is shown in Figure 2.  The pseudo-isothermal fit is shown
as the solid blue line, 
with best-fitting parameters of $V(\infty)=105.4\pm6.1$ km 
s$^{-1}$ and $R_c=1.39\pm0.04$ kpc, and a reduced-$\chi^2$ value of
$\chi^2/\mu=3.19$, where $\mu$ is the degrees of freedom.
The NFW fit is shown as a dotted red line, with best-fitting parameters 
$c_{\rm vir}=4.0\pm1.0$ and $M_{\rm vir}=(2.2\pm0.1)\times10^{11}$ 
M$_{\odot}$, with a 
reduced-$\chi^2$ value of $\chi^2/\mu=1.18$.  As can be seen from Figure
2, the pseudo-isothermal
model (or core model in the figure) 
underestimates the rotation velocities beyond $\sim$7 kpc.  However, the
NFW fit more closely recreates
the observed data.  This is also clear from the values of reduced-$\chi^2$.
We therefore conclude that the NFW model best
represents these data, and this is consistent with the results of
Corbelli \& Salucci (2000).  This is somewhat surprising for a late-type,
bulgeless galaxy like M33, since these late-type galaxies are often
shown to have constant density cores (e.g., Kuzio de Naray, 2006, 2008).

Table 2 lists the best-fit parameters of the 
best-fit NSF and pseudo-isothermal models based upon direct fitting to the 
H{\tt I} rotation curve data. 

\begin{table*}
\label{compare}
\caption{M33 rotation curve modeling results, showing the best-fitting NFW and pseudo-isothermal models.}
\begin{center}
\begin{tabular}{ll}
\hline
Parameter     & NFW model\\
\hline
$M_{\rm vir}$ & $(2.2\pm0.1)\times10^{11}$ M$_{\odot}$\\
$c_{\rm vir}$ & $4.0\pm1.0$\\
$\chi^2/\mu$  & 1.18\\
\hline
Parameter     & Core model\\
\hline
$R_c$            & $1.39\pm0.04$ kpc\\
$V(\infty)$      & $105.4\pm6.1$ km s$^{-1}$\\
$\chi^2/\mu$     & 3.19             \\
\hline
\end{tabular}
\end{center}
\end{table*}

It is probably worthwhile noting that our best-fitting NFW model yields a 
concentration parameter, $c_{\rm vir}=4.0\pm1.0$.  This is somewhat lower than
the concentration parameter of $c_{\rm vir}=5.6$ reported by Corbelli \&
Salucci (2000).  Furthermore, we derive a virial mass of 
$M_{\rm vir}=(2.2\pm0.1)\times10^{11}$ M$_{\odot}$, which is significantly 
lower than the virial mass of $M_{\rm vir}=7.4\times10^{11}$ M$_{\odot}$ 
found by Corbelli \& Salucci (2000).  Here we discuss some reasons that could 
account for these apparent differences.  Since we use the same gas 
distribution as Corbelli \& Salucci (2000), the only difference can come 
from the stellar mass component.  The main difference between our stellar 
mass component, and that of Corbelli \& Salucci (2000), is that ours is 
determined from a Spitzer 3.6-$\mu$m observed in 2007, and that of Corbelli \& 
Salucci (2000) is determined from a $K$ band image reported by (Regan \& 
Vogel 1994).  The $K$ band image from 1994 was taken when near-infrared arrays
were really in their infancy, and so it is probably more important to rely
on the more modern datasets when possible.  Furthermore, Corbelli \& Salucci 
(2000) assume a distance to M33 of 0.7 Mpc, whereas we use the more accurate
measurement of 0.84 Mpc from Magrini et al.\ (2007).  As a result of this
underestimate in the distance to M33, Corbelli \& Salucci (2000) have
underestimated the size of the visible galaxy by a factor of $\sim$17 percent,
and this in turn has probably affected the total mass of M33 that they
derive.  Taking into account the different distances to M33, the disk 
scalelength of $h=1.2$ kpc used by Corbelli \& Salucci (2000) would become 
$h=1.4$ kpc if they had used the more accurate distance of 0.84 Mpc.  This is 
still lower than the scalelngth of $h=1.7$ kpc that we report here.  In 
converting this light distribution into stellar mass, we have then used a 
combination of the stellar
mass-to-light ratios from Bell et al.\ (2003) and the 
population synthesis codes from Maraston (2005).  These papers provide the 
best estimates currently available for determining the stellar mass-to-light 
ratios, and they were not available to Corbelli \& Salucci when they performed 
their analysis. One final difference between our results, and those of 
Corbelli \& Salucci (2000), is that we include the bulge mass, although 
considering the bulge-to-disk ratio of $B/D=0.03$ this is unlikely to have a 
significant effect on the mass models.  As a result, we conclude that the 
differences between our results and those of Corbelli \& Salucci (2000), are 
caused by the different treatment of the disk starlight, updated stellar 
mass-to-light ratios, and more recent data.

Finally, it should be noted that Corbelli \& Walterbos (2007) revealed that
M33 has a weak central bar.  This could potentially have the affect of
inducing non-circular motions in the central regions, i.e., within 1 kpc.
However, Kuzio de Naray \& Kaufmann (2010) have shown that, even in the case
of barred galaxies, it is difficult to confuse an NFW dark matter halo profile 
with that of a pseudo-isothermal profile.  In other words, our result that
M33 is best described by an NFW profile, still holds, and given that the
potential of the stellar bar is weak, the concentration is unlikely to change
significantly.

In the following discussion, we use
the NFW concentration parameter to reveal some interesting relationships.

\section{Discussion}

Seigar et al.\ (2004, 2005, 2006) have demonstrated that a relationship
exists between spiral arm pitch angle and rotation curve shear.  Rotation curve
shear is defined as:
\begin{equation}
S=\frac{A}{\omega}=\frac{1}{2}\left(1-\frac{R}{V}\frac{dV}{dR}\right),
\end{equation}
where $A$ is the first Oort constant, $\omega$ is the angular velocity, and
$V$ is the velocity measured at radius $R$.  Using this equation it is possible
to determine the shear from a rotation curve.  We have performed such an analysis on the H{\tt I} rotation curve for M33 and found a value for its shear of
$S=0.46\pm0.01$. We have also measured the spiral arm pitch angle for M33,
which turns out to be $P=42.^{\circ}2\pm0.^{\circ}3$ (Seigar et al.\ 2008b).
This pitch angle is in good agreement with previous measurements
(Sandage \& Humphreys 1980; Block et al.\ 2004).
Figure 3 shows the relationship between
spiral arm pitch angle and rotation curve shear.  One can easily see that the
pitch angle and shear values for M33 are consistent with the overall 
relationship.

\begin{figure}
\label{shear_vs_pitch}
\includegraphics{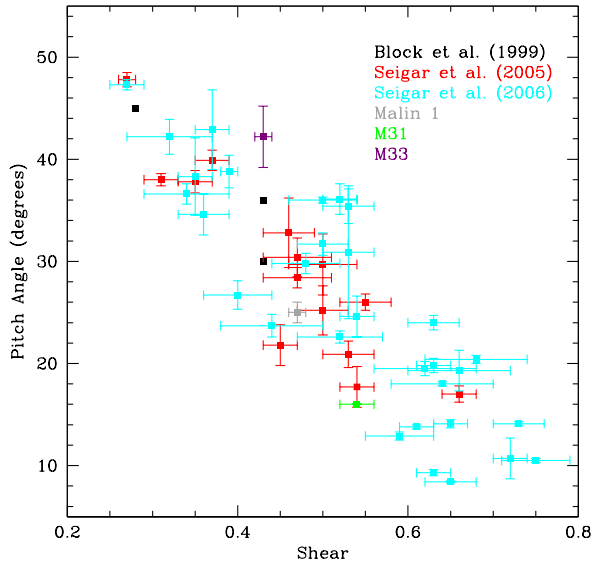}
\vspace*{7cm}
\caption{Spiral arm pitch angle versus rotation curve shear, showing a strong
correlation.  The solid squares represent galaxies with data measured by Block
et al.\ (1999), the red squares are galaxies from Seigar et al.\ (2005), the 
blue squares are galaxies from Seigar et al.\ (2006), the cyan square is for
Malin 1 (Seigar 2008), the green square is for M31 (Seigar et al.\ 2008a) and
the magenta square represents the data for M33 (this paper).}
\end{figure}

Given the spiral arm pitch angles of a number of other galaxies, we can
also now compare this quantity with the NFW concentration parameters for
the galaxies listed in Table 3.  Figure 4
shows a plot of NFW concentration as a function of spiral arm pitch angle
in degrees.  This plot may only be for 5 galaxies, but a relatively strong
correlation appears to exist between these two quantities.  Indeed Pearson's
linear correlation coefficient is 0.95 for this plot, although the significance
at which the null hypothesis of zero correlation is disproved in onlt 54 percent, probably due to low number statistics.  Nevertheless, an interesting 
correlation seems to exist between spiral arm morphology and dark matter concentration, and this could be further studied by targeting more galaxies in an observational campaign.  Indeed, these data seem consistent with the suggestion
that pitch angle and mass concentration are related (Seigar et al.\ 2005, 2006).

\begin{table*}
\label{c_vs_pitch}
\caption{Spiral arm pitch angles, NFW concentration parameters, and central supermassive black hole mass for 5 galaxies.  For the Malin 1 the spiral arm pitch angle is taken from Moore \& Parker (2006).  For M31 the pitch angle is the average of values taken from Arp (1964) and Braun (1991).  The NFW concentration value is taken from (1) Seigar (2008), (2) Klypin et al.\ (2002), (3) Seigar et al.\ (2008a), (4) Seigar et al.\ (2006).  The black hole mass estimates are taken from (5) Ghez et al.\ (2005), (6) Bender et al.\ (2005), (7) Gebhardt et al.\ (2001).}
\begin{center}
\begin{footnotesize}
\begin{tabular}{llllll}
\hline
Galaxy name     &  Spiral arm pitch angle     &  $c_{\rm vir}$   & Source   &  $M_{\rm BH}$     & Source \\
                & (degrees)                   &                  &          & (M$_{\odot}$)\\
\hline
Malin 1         &  $25.0\pm1.0$               & $8.0\pm1.0$      & (1) & -- & -- \\
Milky Way       &  --                         & 12.0             & (2) & $(3.7\pm0.2)\times10^{6}$ & (5) \\
M31             &  $7.1\pm0.4$                & $20.0\pm1.1$     & (3) & $(1.7\pm0.6)\times10^{8}$ & (6)\\
M33             &  $42.2\pm3.0$               & $4.0$            &     & $<1500$      & (7)\\
IC2522          &  $38.8\pm1.6$               & $8.0\pm1.0$      & (4) & -- & -- \\
ESO582G12       &  $22.6\pm0.6$               & $22.0\pm5.0$     & (4) & -- & -- \\
\hline
\end{tabular}
\end{footnotesize}
\end{center}
\end{table*}

\begin{figure}
\label{pitchplot}
\includegraphics{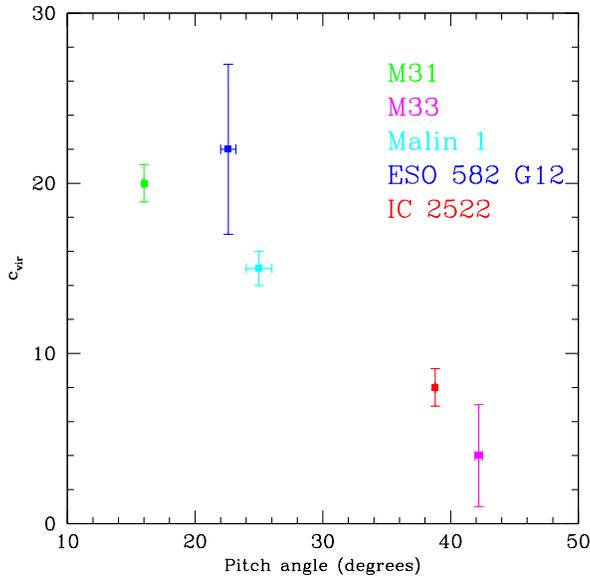}
\vspace*{7cm}
\caption{NFW concentration parameter versus spiral arm pitch angle, showing a correlation.  The green point represents data for M31, the cyan point for Malin 1, the red point for IC2522, the blue point for ESO582G12 and the magenta point shows the data for M33.}
\end{figure}

Finally Figure 5 shows a plot of supermassive black hole mass as a function of 
NFW concentration parameter.  Unfortunately, here we only have data for three 
galaxies.  Nevertheless, a hint of a correlation is starting to show, and 
seeing that such a correlation has been suggested by Seigar et al.\ (2008b), 
as well as Satyapal et al.\ (2008) and Booth \& Schaye (2010), this plot is 
somewhat intriguing.  This hint of a correlation should, of course, be 
expanded on by studying more galaxies along the Hubble sequence from type Sa 
to Sd.

\begin{figure}
\label{bhplot}
\includegraphics{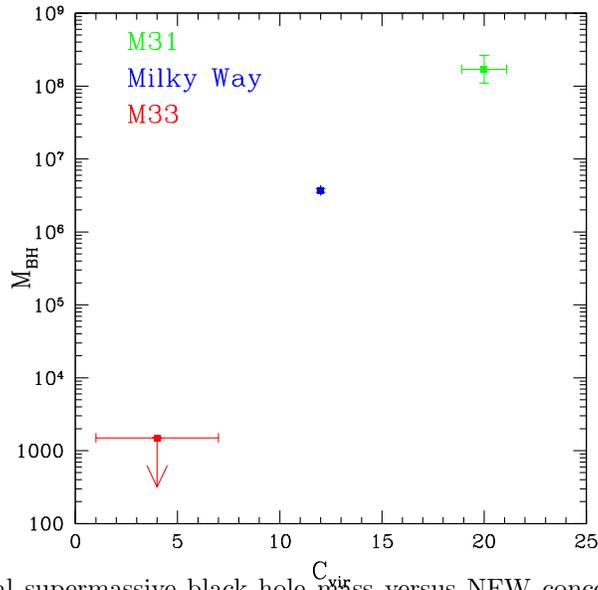}
\vspace*{7cm}
\caption{Central supermassive black hole mass versus NFW concentration parameter, showing a correlation.  The green point represents data for M31, the blue point for the Milky Way and the red point shows the data for M33.}
\end{figure}

\section{Summary}

We have shown that the H{\tt I} rotation curve of M33 can be best modeled with
a dark matter halo that follows a NFW profile, with low NFW concentration of 
$c_{\rm vir}=4.0$.  Using the NFW concentration parameter from this fit, we 
find that interesting correlations between (1) spiral arm pitch angle and NFW 
concentration and (2) central supermassive black hole mass and NFW 
concentration, start to appear.  Although the second correlation is only for
three galaxies, on the surface it appears to be in disagreement with the
argument made by Kormendy \& Bender (2011) that the dark matter halos of 
galaxies have no affect on the masses of supermassive black holes found in 
their centers.  These correlations are very intriguing and our results warrant 
further investigation, as we have been limited to data that was available for 
just a few galaxies.  

\section{Acknowledgements}

This research has made use of the NASA/ IPAC Infrared Science Archive, which is operated by the Jet Propulsion Laboratory, California Institute of Technology, under contract with the National Aeronautics and Space Administration.  The authors wish to thank the reviewers who helped to improve the content of this paper.

%% The Appendices part is started with the command \appendix;
%% appendix sections are then done as normal sections
%% \appendix

%% \section{}
%% \label{}

%% References
%%
%% Following citation commands can be used in the body text:
%% Usage of \cite is as follows:
%%   \cite{key}         ==>>  [#]
%%   \cite[chap. 2]{key} ==>> [#, chap. 2]
%%

%% References with bibTeX database:

%% \bibliographystyle{elsarticle-num}
%% \bibliography{<your-bib-database>}

\begin{thebibliography}{00}

%% \bibitem must have the following form:
%%   \bibitem{key}...
%%

\bibitem{1}Andredakis, Y. C., Peletier, R. F., \& Balcells, M.\ 1995, MNRAS, 275, 874

\bibitem{2}Arp, H.\ 1964, ApJ, 139, 1045

\bibitem{3}Battaglia, G., et al.\ 2005, MNRAS, 364, 433

\bibitem{4}Bell E. F., McIntosh D. H., Katz N., \& Weinberg M. D., 2003, ApJ, 585, 117

\bibitem{5}Bender, R., et al.\ 2005, ApJ, 631, 280

\bibitem{6}Bertin, E., \& Arnouts, S.\ 1996, A\&AS, 117, 393

\bibitem{7}Block, D. L., Puerari, I., Frogel, J. A., Eskridge, P. B., Stockton, A., \& Fuchs, B.\ 1999, Ap\&SS, 269, 5

\bibitem{7a}Block, D.~L., Freeman, K.~C., Jarrett, T.~H., Puerari, I., Worthey, G., Combes, F., \& Groess, R.\ 2004, A\&A, 425, L37

\bibitem{8}Blumenthal, G. R., Faber, S. M., Flores, R., Primack, J. R.\ 1986, ApJ 301, 27

\bibitem{9}Booth, C. M., \& Schaye, J.\ 2010, MNRAS, 405, L1

\bibitem{10}Braun, R.\ 1991, ApJ, 372, 54

\bibitem{11}Bruzual, G., \& Charlot, S., 2003, MNRAS, 344, 1000

\bibitem{12}Bullock, J. S., Dekel, A., Kolatt, T. S., Kravtsov, A. V., Klypin, A. A., Porciani, C., \& Primack, J. R.\ 2001a, ApJ, 555, 240

\bibitem{13}Caon, N., Capaccioli, M., \& D'Onofrio, M.\ 1993, MNRAS, 265, 1013

\bibitem{14}Capaccioli, M.\ 1989, In The World of Galaxies, ed.\ H. G. Corwin, \& L.\ Bottinelli (Berlin: Springer-Verlag), 208

\bibitem{15}Ciotti, L.\ 1991, A\&A, 249, 99

\bibitem{16}Consid\`ere, S., \& Athanassoula, E.\ 1988, A\&AS, 76, 365

\bibitem{17}Corbelli, E., \& Salucci, P.\ 2000, MNRAS, 311, 441

\bibitem{17a}Corbelli, E., \& Walterbos, R.~A.~M.\ 2007, ApJ, 669, 315

\bibitem{18}de Jong, R. S.\ 1996, A\&AS, 118, 557

\bibitem{19}de Vaucouleurs, G.\ 1948, Ann. Astrophys., 11, 247

\bibitem{20}de Vaucouleurs, G.\ 1959, Handbuch der Physik, 53, 275

\bibitem{21}de Vaucouleurs, G., de Vaucouleurs, A., Corwin, H. G., Jr., Buta, R. J., Paturel, G., \& Fouqu\'e, R.\ 1991, The Third Reference Catalog of Bright Galaxies (New York: Springer) (RC3)

\bibitem{22}Diemand, J., Zemp, M., Moore, B., Stadel, J., \& Carollo, C. M.\ 2005, MNRAS 364, 665

\bibitem{23}D'Onofrio, M.\ 2001, MNRAS, 326, 1517

\bibitem{24}Garcia-Gomez, C., \& Athanassoula, E.\ 1993, A\&AS, 100, 431

\bibitem{25}Gebhardt, K., et al.\ 2001, AJ, 122, 2469

\bibitem{26}Gentile, G., Salucci, P., Klein, U., Vergani, D., \& Kalberla, P.\ 2004, MNRAS 351, 903

\bibitem{27}Gentile, G., Burkert, A., Salucci, P., Klein, U., \& Walter, F.\ 2005, ApJ 634, L145

\bibitem{28}Ghez, A. M., Salim, S., Hornstein, S. D., Tanner, A., Lu, J. R., Morris, M., Becklin, E. E., \& Duch\^ene, G.\ 2005, ApJ, 620, 744

\bibitem{29}Graham, A. W.\ 2001a, MNRAS, 326, 543

\bibitem{30}Graham, A. W.\ 2001b, AJ, 121, 820

\bibitem{31}Graham, A. W., \& Driver, S.\ 2005, PASA, 22, 118

\bibitem{32}H\"aring, N., \& Rix, H.-W.\ 2004, ApJ, L89

\bibitem{33}Helou, G., Lu, N. Y., Werner, M. W., Malhotra, S., \& Silbermann, N.\ 2000, ApJ, 532, L21

\bibitem{34}James, P. A., \& Seigar, M. S.\ 1999, A\&A, 350, 791

\bibitem{35}Jedrzejewski, R. I.\ 1987, MNRAS, 226, 747

\bibitem{36}Khosroshahi, H. G., Wadadekar, Y., \& Kembhavi, A.\ 2000, ApJ 533, 162

\bibitem{37}Klypin, A. A., Zhao, H., \& Somerville, R. S.\ 2002, ApJ, 573, 597

\bibitem{37a}Kormendy, J., \& Bender, R.\ 2011, Nature, 469, 377

\bibitem{38}Kuzio de Naray, R., McGaugh, S. S., de Blok, W. J. G., \& Bosma, A.\ 2006, ApJS 165, 461

\bibitem{39}Kuzio de Naray, R., McGaugh, S. S., \& de Blok, W. J. G.\ 2008, ApJ 676, 920

\bibitem{39a}Kuzio de Naray, R., \& Kaufmann, T.\ 2010, MNRAS, submitted (arXiv1012.3471)

\bibitem{40}Magorrian, J., et al.\ 1998, AJ 115, 2285

\bibitem{41}Magrini, L., Corbelli, E., \& Galli, D.\ 2007, A\&A, 470, 843

\bibitem{42}Maraston, C.\ 2005, MNRAS, 362, 799

\bibitem{43}M\"ollenhoff, C., \& Heidt, J.\ 2001, A\&A, 368, 16

\bibitem{44}Moore, L., \& Parker, Q. A.\ 2006, PASA. 23. 165

\bibitem{45}Navarro, J.~F., Frenk, C.~S., \& White, S.~D.~M.\ 1997, ApJ, 490, 493

\bibitem{46}Navarro, J. F., Hayashi, E., Power, C., Jenkins, A. R., Frenk, C. S., White, S. D. M., Springel, V., Stadel, J., \& Quinn, T. R.\ 2004, MNRAS 349, 1039

\bibitem{47}Paturel, G., Petit, C., Prugniel, P., Theureau, G., Rousseau, J., Brouty, M., Dubois, P., \& Cambresy, L.\ 2003, A\&A, 412, 45

\bibitem{48}Prugniel, P., \& Simien, F.\ 1997, A\&A, 321, 111

\bibitem{49}Regan, M.~W., \& Vogel, S.~S., 1994, ApJ, 434, 536

\bibitem{49a}Sandage, A., \& Humphreys, R.~M.\ 1980, ApJ, 236, L1

\bibitem{50}Satyapal, S., Vega, D., Dudik, R. P., Abel, N. P., \& Heckman, T.\ 2008, ApJ, 677, 926

\bibitem{51}Schr\"oder, M. F. S., Pastoriza, M. G., Kepler, S. O., \& Puerari, I., 1994, A\&AS, 108, 41

\bibitem{52}Seigar, M.~S.\ 2008, PASP, 120, 945

\bibitem{53}Seigar, M. S., \& James, P. A.\ 1998, MNRAS, 299, 672

\bibitem{54}Seigar, M. S., Block, D. L., \& Puerari, I.\ 2004, in  Penetrating Bars Through Masks of Cosmic Dust: The Hubble Tuning Fork Strikes a New Note, ed. D. L. Block, I. Puerari, K. C. Freeman, R. Groess, \& E. K. Block (Dordrecht: Springer), 155

\bibitem{55}Seigar, M. S., Block, D. L., Puerari, I., Chorney, N. E., \& James, P. A.\ 2005, MNRAS, 359, 1065

\bibitem{56}Seigar, M. S., Bullock, J. S., Barth, A. J., \& Ho, L. C.\ 2006, ApJ, 645, 1012

\bibitem{57}Seigar, M. S., Barth, A. J., \& Bullock, J. S.\ 2008a, MNRAS, 389, 1911

\bibitem{58}Seigar, M. S., Kennefick, D., Kennefick, J., \& Lacy, C. H. S.\ 2008b, ApJ, 678, L93

\bibitem{59}S\'ersic, J.-L.\ 1963, Bol.\ Asoc.\ Argentina Astron., 6, 41

\bibitem{60}S\'ersic, J.-L.\ 1968, Atlas de Galaxies Australes.\ Observatorio Astronomico, Cordoba

\bibitem{61}Shankar, F., Lapi, A., Salucci, P., de Zotti, G., \& Danese, L.\ 2006, ApJ 643, 14

\bibitem{62}Simon, J.~D., Bolatto, A.~D., Leroy, A., Blitz, L., \& Gates, E.~L.\ 2005, ApJ, 621, 757

\bibitem{63}Spano, M., Marcelin, M., Amram, P., Carignan, C., Epinat, B., \& Hernandez, O.\ 2008, MNRAS, 383, 297

\bibitem{64}Springel, V., et al.\ 2005, Nature 435, 629

\bibitem{65}Swaters, R. A., Madore, B. F., van den Bosch, F. C., \& Balcells, M.\ 2003, ApJ, 583, 732

\bibitem{66}Wechsler, R. H., Bullock, J. S., Primack, J. R., Kravtsov, A. V., \& Dekel, A.\ 2002, ApJ 568, 52

\bibitem{67}Zentner, A. R., \& Bullock, J. S.\ 2002, Phys.\ Rev.\ D, 66, 043003

\end{thebibliography}

%% Authors are advised to submit their bibtex database files. They are
%% requested to list a bibtex style file in the manuscript if they do
%% not want to use elsarticle-num.bst.

%% References without bibTeX database:

\end{document}